# Fehlvorstellungen zur Superposition in der Quantenphysik

Andreas J. C. Woitzik* und Oliver Passon[+]

*Physikalisches Institut, Albert-Ludwigs-Universität Freiburg, Hermann-Herder-Straße 3, 79104 Freiburg im Breisgau
+Bergische Universität Wuppertal Gaußstr. 20, 42119 Wuppertal
Korrespondenz: andreas.woitzik@physik.uni-freiburg.de

**Kurzfassung**

Die Quantenphysik ist seit vielen Jahren ein etablierter Inhalt der Oberstufenphysik. In jüngerer Zeit werden alternative Zugänge über die informationstheoretische Formulierung der Quantentheorie diskutiert, bei denen sich konzeptionelle und begriffliche Schwerpunkte im Vergleich zu etablierten Elementarisierungen verändern. Im Besonderen die Begriffe „Superposition" und „Verschränkung" werden bei diesen Zugängen in das Zentrum der Diskussion gerückt. Ebenfalls besteht ein breites öffentliches Interesse an den neuen Anwendungen der Quantenphysik wie dem Quantencomputer, das zu einer Vielzahl populärer Darstellungen geführt hat. In dieser Arbeit identifizieren wir verbreitete Ungenauigkeiten und Fehler in diesen Darstellungen, erläutern den fachlichen Hintergrund und machen erste Empfehlungen, um Missverständnisse zu vermeiden. Unser Schwerpunkt liegt dabei zunächst auf dem Begriff der „Superposition".

**Abstract**

Quantum physics is a long established content in high school curricula. More recently, alternative approaches based on information-theoretical formulations of quantum theory have been discussed, in which conceptual emphases change in comparison to established didactical approaches. In particular, the notions of "superposition" and "entanglement" are brought into the center of discussion in these approaches. Likewise, there is a broad public interest in the new applications of quantum physics such as quantum computing, which has led to a variety of popular depictions. In this paper we identify common inaccuracies and errors in these representations, explain the technical background and make first recommendations to avoid misunderstandings. Our initial focus is on the notion of "superposition".

## 1. Einleitung

Die Quantenphysik ist seit vielen Jahren ein etablierter Inhalt der Oberstufenphysik. Herkömmliche Lehr- und Bildungspläne folgen dabei grob der historischen Entwicklung und stellen die Energiequantisierung bei gebundenen Zuständen in das Zentrum der Diskussion [1, 2]. Häufig behandelte Inhalte sind folglich das Bohrsche Atommodell, der Franck-Hertz-Versuch oder der lichtelektrische Effekt. Daneben tritt die Behandlung der Interferenz von „Materiewellen", deren Wahrscheinlichkeitsdeutung und schließlich die Unbestimmtheitsrelation. Die analytische Lösung der Schrödinger-Gleichung für den Potenzialtopf stellt oft den mathematischen Höhepunkt der schulischen Behandlung dar. Folgt man dieser Sachstruktur spielt die Superposition von quantenmechanischen Zuständen eine untergeordnete Rolle, die nur selten, wie im Fall der Materiewelleninterferenz (Jönsson-Experiment oder Elektronenbeugung am polykristallinen Grafit) explizit diskutiert wird. Dieser verbreiteten Sachstruktur folgt etwa das Standardlehrwerk *Metzler Physik* [3], indem die Stichworte „Superposition" und „Verschränkung" noch nicht einmal im Schlagwortverzeichnis auftauchen.

In jüngerer Zeit werden jedoch intensiv alternative Zugänge zur Quantenphysik diskutiert, die eine größere Nähe zu aktuell beforschten, technischen Anwendungen herstellen. So sind im Rahmen von QuBIT EDU fachdidaktische Forschungsgruppen vernetzt, die curriculare Entwicklungsarbeit auf dem Gebiet der modernen Quantenphysik (d.h. etwa Quantencomputer und Quantenkryptographie) realisieren wollen.[1] Diese Gruppe zielt dabei nicht nur auf die schulische Bildung, sondern ebenfalls auf hochschuldidaktische Entwicklungen; etwa in den Ingenieurswissenschaften oder der Informatik.

---

[1] http://www.qubit-edu.de/index.php

Im Rahmen eines solchen Ansatzes verschieben sich die konzeptionellen und begrifflichen Schwerpunkte der Darstellung markant. Begriffe wie „Superposition", „Verschränkung" aber auch „Kohärenz" bzw. „Dekohärenz" rücken nun in das Zentrum der Diskussion, während etwa die Diskretheit der Energiezustände keine besonders herausgehobene Rolle mehr spielt. Während zur bisherigen Sachstruktur eine langjährige didaktische Entwicklungsarbeit vorliegt (siehe etwa [4] für einen aktuellen Überblick), hat die Forschung an didaktischen Konzepten zur Quanteninformationstheorie erst begonnen.

Dabei stellt sich eine weitere Herausforderung: Naturgemäß ist die Quanteninformationstheorie an der Grenze zwischen (Quanten-)Physik und Informatik beheimatet. Ihre Behandlung (unabhängig davon, ob als Einführung in die Quantenmechanik oder als aktuelle Anwendung) bezieht sich innerhalb der Quanteninformationstheorie naturgemäß stärker auf Konzepte der Informatik, als dies im üblichen Physikunterricht der Fall ist. Es überrascht daher nicht, dass es ebenfalls Ansätze gibt, das Thema aus der Perspektive der Informatik zu betrachten [5, 6]. Hier liegt ein Potential für fächerübergreifenden Unterricht – aber ebenfalls eine spezielle Herausforderung für die Lehrerinnen- und Lehrerbildung bzw. Fortbildung, da diese Inhalte bisher üblicherweise nicht zum Kanon der Physiklehramtsausbildung gehören (siehe hierzu auch Abschnitt 2.2).

Vor ähnlichen Herausforderungen stehen auch die vielen populären oder semi-populären Darstellungen der Quanteninformationstheorie, die in gedruckter Form oder in online-Quellen die Thematik zu erläutern suchen.

Angesichts des zunehmenden Einsatzes digitaler Quellen (einschließlich reichweitenstarker Erklärvideos) in schulischen und außerschulischen Bildungsprozessen, sehen wir die Notwendigkeit, auch diese Darstellungen zu untersuchen. Der Schwerpunkt unserer Arbeit liegt dabei zunächst auf dem Begriff der „Superposition". Wir untersuchen verbreitete Ungenauigkeiten und Fehler in der Darstellung und machen Empfehlungen für ihre Vermeidung

## 2. Fehlvorstellungen zur Superposition

Das Superpositions- bzw. Überlagerungsprinzip begegnet den Lernenden in zahlreichen Phänomenbereichen der Physik. Bereits in der Mittelstufe lernen die Schülerinnen und Schüler, Kräfte zu addieren (d.h. zu superponieren) und in der Wellenlehre betrachtet man die Überlagerung von Wasser- oder Schallwellen [1, 7]. Obwohl die Quantenmechanik viele Analogien zu anderen Bereichen der Physik aufweist, wird häufig ihre angeblich besondere Merkwürdigkeit betont. So schreibt Beatrice Ellerhoff:

> *„Dieses enorme Rechenpotenzial ist den bizarren Eigenschaften von Quanten zu verdanken. Jedoch erschweren die gleichen Eigenschaften auch das Verstehen der komplexen Vorgänge in Quantencomputern."* [8]

Wir glauben jedoch, dass gerade die Erwähnung von Gemeinsamkeiten mit anderen Phänomenen zur Vermeidung von Fehlvorstellungen und zur Erhöhung der Lernwirksamkeit beitragen kann.

Gleichzeitig gibt es natürlich auch charakteristische Unterschiede zwischen einer Superposition von quantenmechanischen Zustandsvektoren und zum Beispiel Kraftvektoren oder akustischen Schwingungen.

Innerhalb der Quantenphysik besagt das Superpositionsprinzip (zunächst ganz analog zur klassischen Physik), dass die Überlagerung von Zustandsvektoren wieder einen physikalisch zulässigen Zustand beschreibt. Die mathematische Struktur des Zustandsraums in der Quantenmechanik ist ein sogenannter Hilbertraum.[2]

Aber was genau beschreibt ein solcher „Zustandsvektor"? Es handelt sich bekanntlich um eine „Wahrscheinlichkeitsamplitude", d.h. das Quadrat des Zustandsvektors erlaubt (im Gegensatz zur klassischen Physik) lediglich Wahrscheinlichkeitsaussagen über mögliche Messergebnisse. Die genauere Betrachtung der Bedeutung des Zustandsvektors führt dabei rasch in die Debatte um die Interpretation der Quantenmechanik. Allerdings wollen wir diese Diskussion im vorliegenden Artikel ausklammern, obwohl sie natürlich Auswirkungen auf die Vorstellungen und mögliche Fehlvorstellungen hat. Eine Analyse der unterschiedlichen Interpretationen und des Zusammenhangs mit (populär-)wissenschaftlichen Darstellungen könnte womöglich einen fruchtbaren Beitrag zur Debatte bieten, allerdings würde dies den Fokus unserer Arbeit verschieben. Wir beschränken uns hier auf Fehlvorstellungen, die bereits mit dem mathematischen Formalismus nicht vereinbar sind und insofern unabhängig von der Interpretation des Formalismus bedenklich sind.

### 2.1 Die Gleichzeitigkeitsfehlvorstellung

Wählt man eine Basis des Hilbertraumes, so kann man mithilfe dieser Basis jeden seiner Zustände als Überlagerung der Basiszustände darstellen. Für unsere Zwecke beschränken wir uns wie in der Literatur und insbesondere in der populärwissenschaftlichen Literatur üblich auf orthogonale Basissätze und verwenden für ein $n$-Zustands-System die Notation $\{|i\rangle\}$ für die $n$ Basiselemente mit $i$ von *0* bis *n-1*.

---

[2] d.h. ein reeller oder komplexer Vektorraum mit Skalarprodukt, der vollständig bezüglich der durch das Skalarprodukt induzierten Norm ist. In diesem Artikel beschränken wir uns dabei (ebenso wie die von uns untersuchte Literatur) auf endlichdimensionale Hilberträume.

Ein bekanntes Beispiel sind die mit $|+\rangle$ und $|-\rangle$ bezeichneten Superpositionszustände bei gegebenen Basiszuständen $|0\rangle$ und $|1\rangle$ eines Zwei-Niveau-Systems (QuBit):

$$|+\rangle := \frac{(|0\rangle + |1\rangle)}{\sqrt{2}}, \qquad |-\rangle := \frac{(|0\rangle - |1\rangle)}{\sqrt{2}}$$

Häufig werden nun Überlagerungszustände aus den Basiszuständen als solche beschrieben, die *gleichzeitig* diesen und jenen Basiszustand beschreiben. Das folgende Textzitat aus einem online-Medium lautet etwa:

> *„Darüber hinaus kann ein Qubit aber auch beide Zustände GLEICHZEITIG annehmen – und unendlich viele Zustände dazwischen." [9]*

Die Darstellung auf den Webseiten der WDR Wissenschaftssendung *Quarks* (Abbildung 1) verwendet dieselbe Formulierung, indem mehrere Vektoren, die alle jeweils einen Zustand beschreiben, gleichzeitig in eine Bloch-Kugel[3] eingetragen werden.

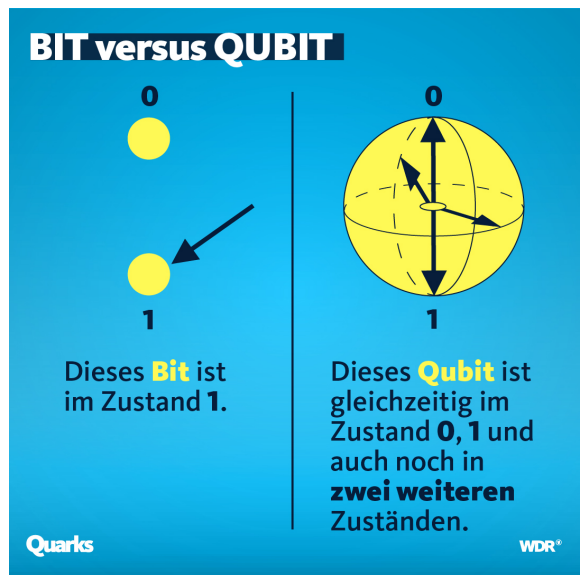

**Abb.1:** Superposition und die Fehlvorstellung der Gleichzeitigkeit.

Der Online-Artikel, der die Abb. 1 enthält, beschreibt den Sachverhalt mit den Worten:

> *„In einem Quantencomputer kann ein Qubit dagegen unendlich viele verschiedene Zustände annehmen und das gleichzeitig." [10]*

Diese Formulierung ist sprachlich unpräzise, denn der Superpositionszustand aus den in Abb. 1 gezeigten Zuständen stellt wieder *einen* Zustand auf der Bloch-Kugel dar. Schließlich führt die Superposition von reinen Zuständen wieder zu *einem* reinen Zustand (und die reinen Zustände befinden sich auf Oberfläche der Bloch-Kugel).

---

[3] Die Bloch-Kugel oder auch Poincaré-Bloch-Kugel ist eine geometrische Darstellung der Zustände eines quantenmechanischen Zwei-Zustandssystems wie zum Beispiel dem Spin eines Elektrons oder der Polarisation eines Photons.

Derselbe Fehler findet sich auch auf dem Kanal *The Morpheus Tutorials [11]* oder sogar in einem Video der renormierten Fraunhofer-Gesellschaft [12]. Dabei werden häufig graphische Element gewählt – wie zum Beispiel das Einzeichnen mehrerer Pfeile in Abb. 1 – die die Gleichzeitigkeitsvorstellung explizit betonen. Selbst die kritische Reflexion über den Begriff wird in der populärwissenschaftlichen Literatur angestellt:

> *„Es ist die Rede davon, dass ein Quantenbit zugleich „0" und „1" ist, oder von Socken, die zugleich rot und blau, aber gleichzeitig einfarbig sind." [13]*

Aus dem oben genannten Grund ist die Gleichzeitigkeitsformulierung jedoch unzutreffend. Dieser Sachverhalt ist in anderen physikalischen Anwendungen des Superpositionsprinzips im Übrigen völlig unstritting. Betrachtet man etwa die Überlagerung von Kräften innerhalb der Mechanik wird stets betont, dass (i) daraus genau *eine* Gesamtkraft resultiert aber ebenfalls, dass (ii) die Komponenten eines Kraftvektors rein konventionell sind, da sie von der Wahl des Koordinatensystems abhängen.

Eine solche Abhängigkeit vom Koordinatensystem (bzw. genauer: von der verwendeten Basis) liegt nun auch in der Quantenmechanik vor und die Gleichzeitigkeitsfehlvorstellung ist dafür zu kritisieren, dass sie diesen Zusammenhang verdeckt. Denn natürlich können auch umgekehrt die Basiszustände $|0\rangle$ und $|1\rangle$ als Superposition der Zustände $|+\rangle$ und $|-\rangle$ dargestellt werden:

$$|0\rangle = \frac{(|+\rangle + |-\rangle)}{\sqrt{2}}, \qquad |1\rangle = \frac{(|+\rangle - |-\rangle)}{\sqrt{2}}.$$

Aus diesen Gründen erscheint es uns günstiger, einen *Superpositionszustand* zu charakterisieren, indem man davon spricht, dass die Komponenten, die superponiert werden (z. Bsp. $|0\rangle$ und $|1\rangle$), nicht *gleichzeitig* vorliegen, sondern als Bestandteile gesehen werden und somit eher *teilweise* vorliegen. Dies kann auch helfen, die oben angesprochene reziproke Beziehung transparent zu machen. Dennoch birgt auch die „teilweise"-Formulierung Gefahr für mögliche Missverständnisse, weil der Begriff aus der Alltagssprache mit verschiedenen Assoziationen verknüpft ist. Auch wird die notwendige Unterscheidung zwischen Amplituden und Wahrscheinlichkeiten von dieser Sprechweise nicht hinreichend umfasst. Es empfiehlt sich daher unserer Meinung nach, auf die unvermeidliche Mehrdeutigkeit unseres Alltagsvokabulars einzugehen und die Problematik explizit zu thematisieren.

### 2.2 Die Parallelitätsfehlvorstellung

Wie in der Einleitung angedeutet, rekurrieren Inhalte der Quanteninformationstheorie naturgemäß stark auf Konzepte der Informatik. In diesem Zusammenhang kann eine Variante der im vorherigen Absatz behandelten „Gleichzeitigkeitsfehlvorstellung" beobachtet werden. Gemeint ist die Behauptung, dass innerhalb eines Quantencomputers viele Berechnungspfade *parallel* (also wiederum „gleichzeitig") bearbeitet werden.

Dies ist jedoch ebenso (bzw. immer noch) unzutreffend und die Diskussion dieser Variante ergänzt die Hinweise des letzten Abschnitts. Das Missverständnis, das dieser Sprechweise vermutlich zugrunde liegt, kann auf verschiedene Arten aufgeklärt werden. Wir geben zunächst eine kurze Diskussion im Rahmen der theoretischen Informatik und betrachten im Anschluss den Messprozess in der Physik genauer.

Eine parallele Bearbeitung der Rechenpfade entspräche komplexitätstheoretisch einer sogenannten „nichtdeterministischen Turingmaschine" (NDTM) [14]. Dieses Konzept der theoretischen Informatik beschreibt ein Maschinenmodell, das in der Lage ist, in jedem Zustand jede mögliche Eingabe gleichzeitig anzunehmen und dann in alle möglichen folgenden Zustände zu wechseln. Die NDTM verallgemeinert damit das Konzept der deterministischen Turingmaschine (DTM), bei der jeder Eingabe ein eindeutig bestimmter Ablauf der Zustände der Maschine entspricht. Die Funktion der NDTM kann in diesem Sinne als Summe von parallel arbeitenden DTM aufgefasst werden. Einfach vorstellen kann man sich den Unterschied einer NDTM und einer DTM beim Ausfüllen eines klassischen 9x9 Sudokus. Während eine DTM ein Feld mit genau einer Zahl ausfüllt und überprüft, ob sich eine erlaubte Lösung ergeben hat, kann die NDTM alle neun Zahlen in dieses Feld eintragen, jede Möglichkeit prüfen und diese Möglichkeiten auch berücksichtigen, wenn sie den folgenden Schritt für das nächste Feld vollführt. Sie arbeitet in diesem Sinne echt parallel.

Diese Konzepte werden innerhalb der theoretischen Informatik verwendet, um sogenannte Komplexitätsklassen zu definieren, also die Frage zu beantworten, welcher Zusammenhang bei einem bestimmten Maschinenmodell zwischen Problemgröße und Rechenzeit besteht.[4] Dabei handelt es sich um ein aktuelles Forschungsfeld mit zahlreichen ungelösten Fragen. Jedoch (und das ist der entscheidende Punkt unseres Arguments) geht man davon aus, dass nichtdeterministischen Turingmaschinen deutlich mächtiger sind als Quantencomputer. [15].

Diese spezielle Parallelitätsfehlvorstellung klingt in der populärwissenschaftlichen Literatur jedoch immer wieder an. So schreibt Ellerhof in ihrem Text aus der *Springer essentials* Reihe (S. 12):

> *„Stattdessen rechnet er [der Quantencomputer] mit allen Qubits zur gleichen Zeit."* [8]

Auch drückt die folgende Formulierung aus einem Erklärvideo eine unglückliche Nähe zum Konzept des Parallelrechners aus:

> *„Aufgrund der Superposition ermöglichen QuBits ganz von allein parallele Rechenoperationen, die sich mit jedem dazukommenden Qubit exponentiell vervielfachen."* [16]

Und schließlich erwähnt zum Beispiel der folgende Artikel im Nachrichtenmagazin *Focus* (zugegebenermaßen keine übertrieben seriöse Quelle zu wissenschaftlichen Fragen, doch dafür reichweitenstark):

> *„Der Quantencomputer rechnet parallel und ist um ein Vielfaches schneller."* [17]

Dieser Fehlvorstellung zu begegnen ist aber auch deshalb so anspruchsvoll, weil der Begriff des „Quantenparallelismus" in einer speziellen technischen Bedeutung sehr wohl etabliert und berechtigt ist. Das Standardwerk zur Quanteninformationstheorie von Nielsen und Chuang weist bereits in seiner 1. Auflage von 2000 auf die Gefahr hin, diesen Begriff falsch zu verstehen:

> *"Quantum parallelism is a fundamental feature of many quantum algorithms. Heuristically, and at the risk of over-simplifying, quantum parallelism allows quantum computers to evaluate a function $f(x)$ for many different values of $x$ simultaneously."* [18]

Man sieht, dass die bisher kritisierten Darstellungen alle an dieser „Übervereinfachung" leiden. Das gedankliche Problem liegt hierbei unter anderem darin, dass die oben beschriebenen Quellen lediglich die Zustandsentwicklung des Rechenregisters betrachten, nicht allerdings die Besonderheiten des quantenmechanischen Mess- und damit Ausleseprozesses. Dieser ist für den Informationsgewinn unumgänglich und spielt eine entscheidende Rolle, wenn Interferenzeffekte in der Quantenmechanik betrachten werden, wie etwa bei der Interferenz von Elektronenstrahlen an Doppelspalt oder polykristallinem Graphit.

Wie kann man sich aber stattdessen erklären, dass Quantencomputer einen Rechenvorteil ermöglichen könnten, und was bedeutet der Begriff des Quantenparallelismus wirklich? Ein wichtiger Unterschied sind die Interferenzeffekte, die die Berechnungen auf einem Quantencomputer von einem klassischen Berechnungspfad auf einem digitalen Computer unterscheiden. Effiziente Algorithmen auf einem Quantencomputer nutzen in der Regel diese Überlagerungen (in einem großen Zustandsraum) geschickt aus – der häufig diskutierte Grover-Algorithmus zur Suche in einer unsortierten Datenbank ist ein Musterbeispiel für diese Strategie [18].

---

[4] So definiert man etwa die Komplexitätsklasse P als die Klasse derjenigen Probleme, bei denen auf einer DTM ein polynomialer Zusammenhang zwischen Problemgröße und Rechenzeit besteht.

Der Fehler der oben beschriebenen Parallelitätsvorstellung kann auch stärker physikalisch begründet werden. Zu diesem Zweck muss man an die Besonderheiten des quantenmechanischen Messprozesses erinnern, denn zu jeder Berechnung einer Problemlösung auf einem Quantencomputer gehört das „Auslesen" des Quantencomputers und dieses entspricht eben einer Messung. Bekanntlich stellen nun die Summanden einer quantenmechanischen Superposition Wahrscheinlichkeitsamplituden dar und jede Messung an einem Superpositionszustand $|\psi\rangle = \sum c_n |n\rangle$ in der entsprechenden Basis führt mit der Wahrscheinlichkeit $|c_n|^2$ auf die jeweilige (und nur *eine*) Komponente der Überlagerung („Bornsche Regel"). Das irreführende Bild der „parallelen" Berechnung zahlreicher Pfade insinuiert jedoch fälschlicherweise, dass ebenfalls zahlreiche („parallele") Ergebnisse gewonnen werden. In seiner terminologischen Bedeutung bezeichnet der Begriff des Quantenparallelismus nun einfach die Tatsache, dass Berechnungen auf Superpositionszuständen erfolgen, aber aus dem oben diskutierten Grund schreiben Nielsen und Chuang in dem bereits zitierten Absatz auch: „However, this parallelism is not immediately useful". Man könnte also vielleicht genauer von einem „pseudo-" oder „quasi-Parallelismus" sprechen, dessen Bedeutung erst dann auftritt, wenn gleichzeitig die Interferenz *zwischen* den betreffenden Überlagerungszuständen zu einem Vorteil genutzt werden kann.

Eine Analogie zu einem Instrument aus der Akustik mag diesen Sachverhalt illustrieren. Bildlich gesprochen müssen verschiedene reine Sinusschwingungen (zum Beispiel Grund- und Obertöne) geschickt miteinander interferiert werden, um den typischen „Klang" eines Instruments zu spielen. Es sei hier angemerkt, dass der Hörprozess weitaus komplizierter ist als hier dargestellt [19], aber für das Anknüpfen an Schulwissen halten wir die Metapher für geeignet. Der Quantencomputer muss am Ende genau den „Klang spielen", der die Lösung einer Rechnung beschreibt. Doch dieser „Klang" kann nicht direkt ausgelesen werden. Die quantenmechanische Messung in einer Basis erlaubt nur mit einer gewissen Wahrscheinlichkeit *einen* Ton (also eine reine Sinusschwingung) auszugeben, der in diesem Klang vorhanden ist. Kurzum: Die Quantenalgorithmen müssen so gestalten sein, dass eine nützliche Information bereits aus (nicht zu vielen solcher) Einzelmessungen gewonnen werden kann.

### 2.3 Die Verwechslung von Superposition und Mischung

Schließlich wollen wir noch auf eine weitere sprachliche Ungenauigkeit aufmerksam machen, die besonders die Anschlussfähigkeit an die (deutsche) Fachsprache beeinträchtigt. Gemeint ist die Unterscheidung bzw. Verwechslung von *Superpositionszuständen* und *gemischten Zuständen*. Auch wenn immer wieder von Zuständen gesprochen wird, die zum Teil aus diesem und zum Teil aus jenem Zustand bestehen, so ist hier doch keine *Mischung* (im terminologischen Sinne) gemeint. In der Quantenmechanik ist strikt zwischen *statistischen Gemischen von Zuständen* und sogenannten *reinen Zuständen* zu unterscheiden. Die Fachliteratur verweist zur Unterscheidung der reinen Zustände und der gemischten Zustände schnell auf den Dichtematrixformalismus nach von Neumann [20, 21]. Dieser hat viele Vorteile, wenn es um die Quantifizierung der Eigenschaften bezüglich Verschränkung oder Mischung geht. Eine Dichtematrix beschreibt danach genau dann einen gemischten Zustand, wenn sie mehrere von Null verschiedene Eigenwerte hat. Bei einem statistischen Gemisch sind verschiedene reine Zustände anwesend und die Wahrscheinlichkeiten drücken bloß die eigene Unkenntnis aus.

In der populärwissenschaftlichen Literatur wird nun häufig eine Sprache verwendet, die der Fachterminologie widerspricht. Damit ist das vermittelte Wissen nicht anschlussfähig.

Beispiele für dieses Problem sind die Formulierungen: „Zustände, die keine Basiszustände sind, heißen „gemischte Zustände"." [13] oder auch „Das Hadamard-Gatter hat hierbei die Aufgabe den Zustand eines Qubits von einer reinen „Null" oder „Eins" in eine Mischung, welche zu gleichen Teilen aus „Null" und „Eins" besteht, zu überführen." [8]

Die Unterscheidung von reinen und gemischten Zuständen bereitet Studierenden bereits ohne etablierte Fehlvorstellungen Schwierigkeiten. Begegnen Lernende diesen problematischen Formulierungen bereits im schulischen Kontext, dürfte dies ein weiteres Hemmnis darstellen.

### 3. Zusammenfassung und Ausblick

Unserer Ansicht nach sind im Zusammenhang mit (nicht nur) quantenmechanischen Superpositionszuständen die Begriffe der „Gleichzeitigkeit" oder „Parallelität" aus den oben genannten Gründen mit besonderer Vorsicht zu verwenden. In der Fachsprache bereits eindeutig verwendete Begriffe wie die „Mischung" sollten nur im richtigen Zusammenhang einer klassischen Mischung verwendet und nicht mit der Superposition zweier Zustände verwechselt werden.

Dazu kann man zum Beispiel wieder den Vergleich zu Schallwellen oder Kräften als bekanntes Bild bemühen, auch wenn – und das ist sicherlich ein großer Nachteil – hier das Analogon zur quantenmechanischen Messung fehlt. Betrachtet man etwa innerhalb der Mechanik eine Kraft als Überlagerung mehrerer Teilkräfte ($F_{ges} = \sum F_i$) ist es weder üblich noch sinnvoll, davon zu sprechen, dass alle $F_i$ „gleichzeitig" vorlägen. Vielmehr hat man es mit *einer* Gesamtkraft zu tun, die sie aus der Superposition der Teilkräfte ergibt. Wir halten daher in diesem Kontext den Begriff *teilweise* gegenüber *gleichzeitig* für angemessener. Eine andere Möglichkeit ist die Verwendung von Metaphern aus dem Alltag, wie sie in [22] vorgeschlagen wird. Hier werden Superpositionszustände als Entscheidungszustände in der Schwebe beschrieben, zum Beispiel, wenn noch nicht entschieden ist, ob das Mittagessen im Restaurant A oder B gegessen werden solle.

Bereits in früheren didaktischen Arbeiten wurde auf die Notwendigkeit der fachlichen Korrektheit und Anschlussfähigkeit der Konzepte der Quantenphysik hingewiesen [23]. Darüber hinaus ist die fachliche Korrektheit notwendig, um den Lernenden und der Gesellschaft eine fundierte Bewertung der Möglichkeiten von Quantentechnologien, bzw. der Nutzung quantenphysikalischer Prinzipien, zu ermöglichen.

Gerade bei aktuellen Themen gewinnen außerdem populärwissenschaftliche Darstellungen (sei es im Internet auf gängigen Videoportalen oder in der Literatur) an Bedeutung. Dabei darf man annehmen, dass diese sowohl von Lehrerinnen und Lehrern zur Unterrichtsvorbereitung genutzt, als auch von besonders motivierten Lernenden konsumiert werden.

Ein besonderes Problem der populärwissenschaftlichen Quellen ist jedoch die fehlende Qualitätssicherung. So haben wir beispielsweise bei durchaus namhaften Verlagen und wissenschaftlichen Gesellschaften Fehlvorstellungen auf den Kommunikationskanälen für die Öffentlichkeit entdeckt. Auf Hinweise unsererseits wurde nur zaghaft oder gar nicht reagiert. Der beobachtbare *Hype* der Quanteninformationsverarbeitung und mit ihr einhergehender Technologien kann eine unrealistisch hohe Erwartungshaltung der Gesellschaft bewirken, wenn falsche oder ungenaue Vorstellungen kommuniziert werden. Die Erwartungen können letztendlich nicht erfüllt werden und führen im schlimmsten Fall zu einem Ansehensverlust der entsprechenden wissenschaftlichen Felder und ihrer Vertreterinnen und Vertreter.

### 4. Literatur nach DIN 1505 oder APA (ohne [x])

[1] Ministerium für Kultus, Jugend und Sport Baden-Württemberg: Bildungsplan 2016, Gymnasium – Physik – Überarbeitete Fassung vom 25.03.2022 (V2), 2022. URL: https://www.bildungsplaene-bw.de/,Lde/LS/BP2016BW/ALLG/GYM/PH.V2 (Stand 02.09.2022)

[2] Staatsministerium für Schulqualität und Bildungsforschung. Lehrplan Physik Gymnasium Bayern, Jahrgangsstufe 12, 2012. URL: https://www.isb.bayern.de/gymnasium/lehrplan/gymnasium/jahrgangsstufenprofile-ebene-3/12-jahrgangsstufe/518/ (Stand 02.09.2022)

[3] Grehn, Joachim; Krause, Joachim (2015): Metzler Physik Gesamtband SII. Grundkurs NRW, Braunschweig: Bildungshaus Schulbuchverlage.

[4] Wilhelm, T.; Schecker, H.; Hopf, M. (2021) Unterrichtskonzeptionen für den Physikunterricht. Heidelberg: Springer.

[5] Woitzik, Andreas (2020): Quanteninformationsverarbeitung in der gymnasialen Oberstufe. In LOG IN: Vol 41, No. 1. Berlin: LOG IN Verlag, S. 39-46

[6] Michaeli, Tilman; Seegerer, Stefan; Romeike Rolf (2021): Quanteninformatik als Thema und Aufgabengebiet informatischer Bildung. In: Informatik – Bildung von Lehrkräften in allen Phasen, Lecture Notes in Informatics (LNI) 2021, S. 123-132

[7] KMK (2020): Bildungsstandards im Fach Physik für die Allgemeine Hochschulreife: URL: https://www.kmk.org/fileadmin/Dateien/veroeffentlichungen_beschluesse/2020/2020_06_18-BildungsstandardsAHR_Physik.pdf (Stand 02.09.2022)

[8] Ellerhoff, Beatrice Marie (2020): Mit Quanten Rechnen, Wiesbaden: Springer Spektrum.

[9] Homepage Vodafone: URL: https://www.vodafone.de/business/featured/technologie/quantencomputer-so-funktioniert-er-das-kann-er/ (Stand 12.09.2022)

[10] Homepage Quarks: URL: https://www.quarks.de/technik/faq-so-funktioniert-ein-quantencomputer/ (Stand 25.03.2022)

[11] YouTube-Video des Kanals The Morpheus Tutorials: URL: https://www.youtube.com/watch?v=kBLYe6_IgPs&t=260s

[12] YouTube-Video der Fraunhofer Gesellschaft: URL: https://www.youtube.com/watch?v=m67jr1KQES0&t=104s

[13] Just, Bettina (2020): Quantencomputing kompakt, Berlin, Heidelberg: Springer Vieweg.

[14] Hopfcroft, John E.; Motwani, Rajeev; Ullman, Jeffrey D. (2014): Introduction to automata theory, languages, and computation, Harlow: Pearson Education.

[15] Aaronson, Scott (2010): BQP and the polynomial hierarchy. In: STOC '10: Proceedings of the forty-second ACM symposium on Theory of computing 2010, 141-150 June 2010 Pages 141–150 https://doi.org/10.1145/1806689.1806711

[16] YouTube-Video des Kanals brainfaqk: URL: https://www.youtube.com/watch?v=u5S-hEvq9SI&t=260s (Stand: 12.09.22)

[17] Online-Artikel auf Focus.de: URL: https://www.focus.de/finanzen/boerse/aktien/tiefgekuehlter-superrechner-steuert-die-welt-von-morgen_id_13435226.html (Stand: 28.03.2022)

[18] Nielsen, Michael A.; Chuang, Isaac L. (2002): Quantum computation and quantum information, Cambridge: Cambridge University Press.

[19] Kern, Albert; Stoop, Ruedi: Essential Role of Couplings between Hearing Nonlinearities. In:

Physical Review Letters Volume 91, Nummer 12 (2003).
[20] Von Neumann, John (1927): Wahrscheinlichkeitstheoretischer Aufbau der Quantenmechanik. In: Nachrichten von der Gesellschaft der Wissenschaften zu Göttingen, Mathematik-Physikalische Klasse 1927, S. 245-272.
[21] Cohen-Tannoudji, Claude; Diu, Bernard; Laloë Franck (1986): Quantum Mechanics Volume 1, Paris: Hermann.
[22] Pospiech, Gesche (2021): Quantencomputer & Co. Wiesbaden: Springer Spektrum.
[23] Pospiech, Gesche; Schöne, Matthias: Quantenphysik in Schule und Hochschule. In: PhyDid B-Didaktik der Physik -Beiträge zur DPG-Frühjahrstagung 2012.

**Danksagung**